\documentclass[journal]{IEEEtran}

\usepackage{epsfig,amsmath,amssymb,epsf,amsthm,scalefnt,multirow}
\usepackage{xcolor}
\usepackage{float}
\usepackage{cite}
\newcommand{\argmax}{\mathop{\mathrm{argmax}}}
\newcommand{\argmin}{\mathop{\mathrm{argmin}}}

  \def\cC{{\mathcal{C}}}

 \def\cN{{\mathcal{N}}}

\def\argmin{\mathop{\mathrm{argmin}}}
\def\argmax{\mathop{\mathrm{argmax}}}
\def\diag{\mathop{\mathrm{diag}}}

\def\b0{{\pmb{0}}} 

\def\ba{{\mathbf{a}}} \def\bb{{\mathbf{b}}} \def\bc{{\mathbf{c}}} 
 \def\bff{{\mathbf{f}}} \def\bg{{\mathbf{g}}} \def\bh{{\mathbf{h}}}
   
   \def\bp{{\mathbf{p}}}
   
\def\bu{{\mathbf{u}}}   \def\bx{{\mathbf{x}}}
\def\by{{\mathbf{y}}}   

\def\bA{{\mathbf{A}}}  \def\bC{{\mathbf{C}}} 
 \def\bF{{\mathbf{F}}}  \def\bH{{\mathbf{H}}}
\def\bI{{\mathbf{I}}}   
   \def\bP{{\mathbf{P}}}
 \def\bR{{\mathbf{R}}}  
\def\bU{{\mathbf{U}}}

\begin{document}

\title{Trellis-Extended Codebooks and Successive Phase Adjustment: A Path from LTE-Advanced to FDD Massive MIMO Systems}

\author{Junil Choi, David J. Love, and Taeyoung Kim\\
\thanks{Junil Choi and David J. Love are with the School of Electrical and Computer Engineering, Purdue University, West Lafayette, IN (e-mail: choi215@purdue.edu, djlove@purdue.edu).}
\thanks{Taeyoung Kim is with the Advanced Communications Lab., Communications Research Team, DMC R\&D Center, Samsung Electronics Co., Ltd. (e-mail: ty33.kim@samsung.com).}}


\maketitle

\begin{abstract}
It is of great interest to develop efficient ways to acquire accurate channel state information (CSI) for frequency division duplexing (FDD) massive multiple-input multiple-output (MIMO) systems that achieve backward compatibility.  It is theoretically well known that the codebook size in bits for CSI quantization should be increased as the number of transmit antennas becomes larger, and 3GPP Long Term Evolution (LTE) and LTE-Advanced codebooks also follow this trend.  Thus, in massive MIMO, it is hard to apply the conventional approach of using pre-defined vector-quantized codebooks for CSI quantization because of codeword search complexity.  In this paper, we propose a trellis-extended codebook (TEC) that can be easily harmonized with current wireless standards such as LTE or LTE-Advanced by extending standardized codebooks designed for two, four, or eight antennas with trellis structures.  TEC exploits a Viterbi decoder and a convolutional encoder as a CSI quantizer and a CSI reconstructor, respectively.  By quantizing multiple channel entries simultaneously using standardized codebooks in a state transition of a trellis search, TEC can achieve fractional number of bits per channel entry quantization and a practical feedback overhead.  Thus, TEC can solve both the complexity and the feedback overhead issues of CSI quantization in massive MIMO systems.  We also develop trellis-extended successive phase adjustment (TE-SPA) which works as a differential codebook for TEC.  This is similar to the dual codebook concept of LTE-Advanced.  TE-SPA can reduce CSI quantization error with lower feedback overhead in temporally and spatially correlated channels.  Numerical results verify the effectiveness of the proposed schemes in FDD massive MIMO systems.
\end{abstract}


\section{Introduction}\label{sec1}
The 3GPP LTE-Advanced standard has been recently finalized \cite{lte_advanced}, and commercial products that support LTE-Advanced are about to be released worldwide.  LTE-Advanced is able to deploy up to eight antennas at the base station meaning that efficient downlink and uplink multiple-input multiple-output (MIMO) techniques can be exploited \cite{lte_advanced_mimo}.  To quantize channel state information (CSI) more efficiently, LTE-Advanced introduced dual codebooks that quantize long-term/wideband and short-term/subband CSI separately \cite{LTE_A_cb}.  However, LTE-Advanced still relies on pre-defined vector-quantized codebooks with a codeword search complexity that grows exponentially with the codebook size.

Recently, the idea of using a very large number of antennas at the base station, known as massive MIMO or large-scale MIMO, has drawn considerable interest from both academia and industry to further enhance the total network throughput by enabling aggressive multi-user MIMO (MU-MIMO) systems \cite{massive_mimo5,fdmimo}.  To exploit the full benefits of massive MIMO, the base station needs to have accurate CSI for all scheduled downlink channels.  Thus, the challenge is to scale channel estimation and feedback methods to provide CSI effectively.  Most of the literature on massive MIMO focuses on time division duplexing (TDD) to sidestep this challenge.  By relying on TDD, CSI can be extracted implicitly by using uplink pilot signals and the downlink/uplink channel reciprocity property assuming the transmit and receive antennas RF chains are properly calibrated \cite{tdd_cali}.  However, frequency division duplexing (FDD) dominates most of the cellular market today, and it is expected that FDD would be adopted for at least the first stage of massive MIMO deployments when the number of transmit antennas is \textit{not very large}, e.g., 32 or 64 antennas, for backward compatibility \cite{fdmimo}.  Thus, it is of great interest to develop efficient ways to acquire accurate CSI for FDD massive MIMO systems.

To implement FDD massive MIMO systems, we need to develop 1) a novel training technique for downlink channel estimation and 2) an efficient CSI quantization method.  Note that the overhead of downlink training (relying on conventional unitary training techniques) and CSI quantization must both scale proportional to the number of transmit antennas to enable accurate channel estimation at the user and to maintain a certain level of CSI quantization loss \cite{training1,rvq3}.  Because of the very large number of antennas, the overhead for both unitary training and vector-quantized codebook based CSI feedback might overwhelm the downlink and uplink resources in massive MIMO systems.  Moreover, the complexity of CSI quantization using vector-quantized codebooks increases exponentially with the feedback overhead (or the number of transmit antennas).  The heavy training/feedback overhead and CSI quantization complexity problems should be solved to implement practical FDD massive MIMO systems.

Recently, several works have been dedicated to solving these issues. In \cite{closedloop_training,cl_training_jstsp,song}, efficient training techniques with temporal overheads that do not increase linearly with the number of transmit antennas have been proposed.  For the CSI quantization issue, a compressed sensing based CSI quantization approach is proposed in \cite{kuo:2012}, and an overall feedback reduction technique that exploits spatial correlation of users is proposed in \cite{nam:2012}.

In this paper, we propose a trellis-extended codebook (TEC) for FDD massive MIMO systems.  TEC combines a trellis encoder (i.e., Viterbi decoding) and a vector-quantized codebook to quantize a large dimensional channel.  Similar CSI quantization approaches that exploit the trellis encoder have been proposed in \cite{CK,tcom_ntcq}.  In \cite{CK}, a trellis search and vector-quantized codebook are both considered as in TEC.  However, the path metric for the trellis search is hand-optimized in \cite{CK}, resulting in severe performance degradation.  The duality between noncoherent sequence detection and CSI quantization is exploited in \cite{tcom_ntcq} to set an appropriate path metric for the trellis search.  However, \cite{tcom_ntcq} relies on standard constellation points such as phase shift keying (PSK) or quadrature amplitude modulation (QAM) to quantize a channel vector, which gives a minimum feedback overhead of one bit per channel entry and does not have a straightforward extension of existing 3GPP codebooks.

The proposed TEC adopts the same path metric as in \cite{tcom_ntcq}, but TEC utilizes vector-quantized codebooks rather than constellation points.  Therefore, TEC can easily satisfy backward compatibility by exploiting standardized LTE or LTE-Advanced codebooks\footnote{Instead of having different CSI quantization methods for different number of transmit antennas, it is desirable to reuse standardized feedback frameworks, e.g., LTE or LTE-Advanced codebooks, in practice.} and achieve a fractional number of bits per channel entry quantization to allow practical feedback overhead.  TEC can utilize other codebooks, e.g., a Grassmannian line packing (GLP) codebook \cite{grass1,grass3}, a random vector quantized (RVQ) codebook \cite{rvq1,rvq3}.  We develop a codeword-to-branch mapping rule to maximize the performance of TEC.  The numerical results show that the mapping rule gives a non-negligible gain over TEC even with the same codebook. We also investigate a codebook design methodology (instead of reusing conventional codebooks) that is suitable to TEC in this paper.

Moreover, we propose trellis-extended successive phase adjustment (TE-SPA) which functions as a differential version of TEC.  Note that differential codebooks exploit temporal correlation of channels to reduce quantization error in a successive manner \cite{tm_correlated1,tm_correlated2,tm_correlated7,tm_correlated8,tm_correlated3,tm_correlated4,tm_correlated9,tm_correlated5,tm_correlated6}.  TE-SPA quantizes channels successively in time and can reduce quantization loss even with a reduced feedback overhead than TEC.  We show that TE-SPA can be applied to spatially correlated channels as well without any changes. The concept of TEC and TE-SPA is similar to LTE-Advanced dual codebooks, i.e., TEC quantizes long-term/wideband CSI while TE-SPA quantizes short-term/subband CSI.  This unified structure for long-term/wideband and short-term/subband CSI quantization is a significant benefit compared to other stand-alone CSI quantization schemes for massive MIMO systems.

The remainder of this paper is organized as follows.  We explain the system model we consider in Section \ref{sec2}. The proposed TEC and TE-SPA are explained in Section \ref{sec3} and Section \ref{sec4}, respectively.  Simulation results are presented in Section \ref{sec5}, and conclusions follow in Section \ref{sec6}.

\section{System Model}\label{sec2}
\begin{figure}[t]
  \centering
  \includegraphics[scale = 0.45]{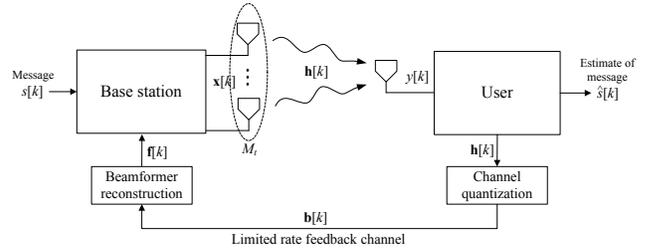}\\
  \caption{Multiple-input single-output communications system with limited feedback.}\label{MISO}
\end{figure}
To simplify explanation, we first consider a block fading multiple-input single-output (MISO) channel with $M_t$ transmit antennas at the base station and a single receive antenna at the user as shown in Fig. \ref{MISO}.  The proposed TEC can be easily extended to a multiple receive antenna case as explained in Section \ref{TEC_MIMO}.  With the block fading assumption, the received signal in the $k$th fading block, $y[k] \in \mathbb{C}$, is written as
\begin{equation*}
y[k] = \sqrt{P}\bh^H[k]\bff[k]s[k] + z[k],
\end{equation*}
where $P$ is the transmit power, $\bh[k] \in \mathbb{C}^{M_t}$ is the MISO channel vector, $\bff[k] \in \mathbb{C}^{M_t}$ is the unit norm beamforming vector, $s[k] \in \mathbb{C}$ is the message signal satisfying $E\left[s\left[k\right]\right] = 0$ and $E\left[|s[k]|^2\right] = 1$, and $z[k] \sim \cC\cN(0,\sigma^2)$ is complex additive white Gaussian noise.

For CSI quantization, we assume that the total number of feedback bits $B_{tot}$ scales linearly with $M_t$ as
\begin{equation*}
    B_{tot} \triangleq B M_t
\end{equation*}
where $B$ is the number of quantization bits per transmit antenna.  The linear increment of the feedback overhead is necessary to achieve a certain level of channel quantization error \cite{rvq3} or a full multiplexing gain of MU-MIMO \cite{mumimo1,mumimo2}.

If we rely on the conventional approach of using a $B_{tot}$-bit unstructured vector-quantized codebook $\cC = \{\bc_{1},\ldots,\bc_{2^{B_{tot}}}\}$ that consists of unit norm codewords for CSI quantization, the user quantizes its channel by selecting the best codeword $\bc_{\mathrm{opt}}[k]$ that aligns with the channel most closely as
\begin{equation} \label{cb_select}
\bc_{\mathrm{opt}}[k] = \argmax_{\bc \in \cC} |\bh^H[k]\bc|^2.
\end{equation}
The user then feeds back the binary index of $\bc_{\mathrm{opt}}[k]$, i.e., $\bb[k]=\mathrm{bin}(\mathrm{opt})$ where $\mathrm{bin}(\cdot)$ converts an integer to its binary representation, to the base station.  If the base station adopts maximum ratio transmission (MRT) beamforming, which is popular due to its simplicity for massive MIMO \cite{massive_mimo5}, we have $\bff[k]=\bc_{\mathrm{opt}}[k]$.

Note that the codeword search complexity of using a vector-quantized codebook is $O(M_t 2^{B M_t})$.  If $B_{tot}$ or $M_t$ is small as in current cellular systems, the complexity of CSI quantization is not a problem.  However, in massive MIMO systems with a very large number of $M_t$, brute force codeword selection becomes infeasible.

\section{Trellis-Extended Codebook (TEC)}\label{sec3}
TEC can exploit and extend pre-existing vector-quantized codebooks such as LTE or LTE-Advanced codebooks.  Because of its backward compatibility, TEC is an excellent candidate for CSI quantization in future FDD massive MIMO systems.  We first explain the concept and the procedure of TEC.  We then discuss the codeword-to-branch mapping and codebook design criteria to maximize the performance of TEC.  Because we do not consider temporal correlation of channels in this section, we drop the block index $k$ to simplify notations for the remainder of this section.

\subsection{Concept and procedure of TEC}
Similar to \cite{CK,tcom_ntcq}, TEC exploits a trellis decoder and a convolutional encoder in channel coding as a CSI quantizer and a CSI reconstructor, respectively.  Vector-quantized codewords (e.g., codebooks designed for smaller arrays) are mapped to trellis branches to quantize multiple channel entries simultaneously by the Viterbi algorithm.  We first explain the concept of TEC in detail.  Then, we summarize the procedure of TEC.

Like \cite{tcom_ntcq}, TEC is based on the equivalence between the two optimization problems
\begin{equation*}
\hat{\bx} = \argmin_{\bx\in \mathbb{C}^{N}}\min_{\theta\in [0,2\pi)}\left\Vert\by - e^{j\theta}\frac{\bx}{\|\bx\|_2}\right\Vert_{2}^2
\end{equation*}
and
\begin{equation}
\hat{\bx} = \argmax_{\bx\in \mathbb{C}^{N}}\frac{|\by^H\bx|^2}{\left\Vert\bx\right\Vert_2^2}.\label{equiv1}
\end{equation}
Note that \eqref{equiv1} is the same as \eqref{cb_select}.  Thus, with the constraint of $\|\bc\|_2^2=1$, we can transform the CSI quantization problem in \eqref{cb_select} to
\begin{equation}
\bc_{\mathrm{opt}} = \argmin_{\bc\in \cC}\min_{\theta\in [0,2\pi)}\left\Vert\bh - e^{j\theta}\bc\right\Vert_{2}^2.\label{ntcq_cb_select}
\end{equation}
Instead of optimizing $\theta$ over the continuous space $[0,2\pi)$, we can discretize the search space, i.e., $\theta\in \Theta=\left\{\theta_1,\ldots,\theta_{K_{\theta}}\right\}$, as in noncoherent sequence detection \cite{Madhow}.  With a given $\theta$, \eqref{ntcq_cb_select} can be efficiently solved by well-known source coding techniques such as trellis coded modulation (TCQ) or trellis quantizer \cite{tcq1}.  This conversion is successfully exploited in \cite{QAM,tcom_ntcq} to develop efficient CSI quantizers.  TEC also solves \eqref{ntcq_cb_select} using trellis quantizers similar to \cite{tcom_ntcq}.  The main difference is that \cite{tcom_ntcq} handles one channel entry per state transition of the trellis search while TEC processes multiple channel entries simultaneously.

\begin{figure}[t]
  \centering
  \includegraphics[scale = 0.5]{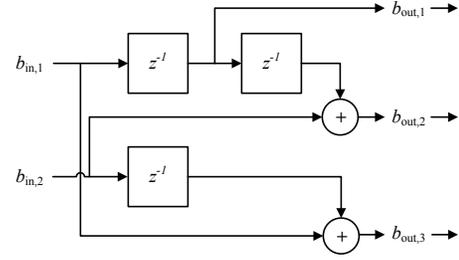}\\
  \caption{A rate $\frac{2}{3}$ convolutional encoder that can be used to generate a TEC codebook. In the figure, $b_{\mathrm{in},1}$ and $b_{\mathrm{in},2}$ are the least significant and the most significant input bits, respectively.  Same for the output bits.}\label{conv_enc}
\end{figure}

\begin{figure}[t]
  \centering
  \includegraphics[scale = 0.65]{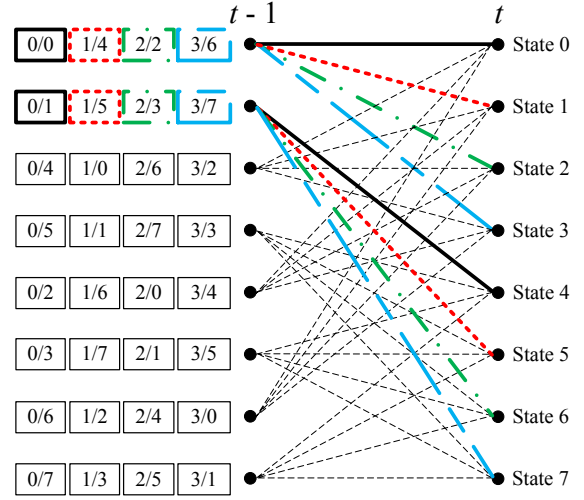}\\
  \caption{The trellis representation of the convolutional encoder in Fig. \ref{conv_enc}.  Each state transition in the right side is mapped with input/output relation using decimal numbers in each box in the left. For example, 1/4 (in decimal numbers) in the top red-dot box represents the state transition from the state 0 to the state 1 with input=01/output=100 (all in binary numbers).}\label{trellis_rep}
\end{figure}

TEC can be implemented using any trellis quantizer.  In this paper, we adopt the Ungerboeck trellis and convolutional encoder \cite{Unger} because of their simplicity and good performance.  Let $B_{in}$ and $B_{out}$ be the number of input and output bits of a convolutional encoder of interest, respectively.  The Ungerboeck convolutional encoder satisfies $B_{out}=B_{in}+1$.  Note that each state in the trellis of the corresponding convolutional encoder has $2^{B_{in}}$ branches; however, the total number of distinctive branches is $2^{B_{out}}$.  An example of a rate $\frac{2}{3}$ convolutional encoder from \cite{Unger} and the corresponding trellis are shown in Fig. \ref{conv_enc} and \ref{trellis_rep}, respectively.  As shown in Fig.~\ref{trellis_rep}, each state has four branches differentiated with inputs and even or odd outputs.

Let $L$ denote the number of simultaneously quantized channel elements in a state transition of a trellis.  We assume that $L$ divides the number of transmit antennas $M_t$.  Note that TEC supports $B=\frac{B_{in}}{L}$ bits per channel entry quantization, which will become clear later.  Thus, if $L>B_{in}$, TEC can achieve a fractional number of bits per channel entry quantization.

To process $L$ channel entries per state transition, TEC maps $L \times 1$ codewords $\bc^{L}_{k}\in \mathbb{C}^{L}$ to branches in the trellis.  To do this, we need to have a vector-quantized codebook (such as the LTE codebook) with $2^{B_{out}}$ codewords, i.e., $\mathcal{C}^L_{2^{B_{out}}}=\left\{\bc_1^L,\ldots,\bc_{2^{B_{out}}}^L\right\}$, to assign all $2^{B_{out}}$ branches of the trellis with different output.  We will discuss the codeword-to-branch (or outputs) mapping and the codebook design criteria later.  For the time being, we assume that all $2^{B_{out}}$ branches are mapped with some codewords.

To perform the trellis search using the Viterbi algorithm, we need to define a path metric to solve \eqref{ntcq_cb_select}.  Let $\bp_t$ be a partial path up to the stage $t$ in the trellis.  We also define $\mathrm{in}(\bp_t)$ as the binary input sequence corresponding to path $\bp_t$ and $\mathrm{out}(\bp_t)$ as the sequence of codewords $\bc^{L}_{k}$'s that are mapped to branches in the path $\bp_t$.  Note that $\mathrm{out}(\bp_t)\in \mathbb{C}^{Lt}$ where each block of $L$ entries of $\mathrm{out}(\bp_t)$ is from a specific codeword $\bc^{L}_{k}$.  With these definitions, we can define the path metric based on \eqref{ntcq_cb_select} as
\begin{align}
& \nonumber m(\bp_t,\theta)\\
&\nonumber ~ = \left\Vert\bh_{[1:Lt]}-e^{j\theta}\mathrm{out}(\bp_t)\right\Vert_2^2\\
&~ = m(\bp_{t-1},\theta) + \left\Vert\bh_{[L(t-1)+1:Lt]}-e^{j\theta}\mathrm{out}([p_{t-1}~p_t])\right\Vert_2^2 \label{TEC_path_metric}
\end{align}
where $\bh_{[m:n]}$ is the truncated vector of $\bh$ from the $m$th entry to the $n$th entry.  The path metric in \eqref{TEC_path_metric} can be efficiently computed for a given candidate value of $\theta$ using the Viterbi algorithm where the total number of stages in the trellis is equal to $T=\frac{M_t}{L}$.  The best path $\bp_{\mathrm{best}}$ and the best phase $\theta_{\mathrm{best}}$ that minimize the path metric in \eqref{TEC_path_metric} are given by solving
\begin{equation*}
\min_{\theta\in \Theta}\min_{\bp_{T}\in\mathbb{P}_{T}}m(\bp_{T},\theta)\label{pbest}
\end{equation*}
where $\mathbb{P}_{T}$ denotes the set of all possible paths up to stage $T$.  The best codeword $\bc_{\mathrm{opt}}$ and the binary feedback sequence $\bb$ are given as
\begin{equation}\label{c_and_b}
\bc_{\mathrm{opt}}=\mathrm{out}(\bp_{\mathrm{best}}),~~\bb=\mathrm{in}(\bp_{\mathrm{best}}),
\end{equation}
respectively.  If we normalize $\bc^{L}_{k}$ as $\|\bc^{L}_{k}\|_2^2=\frac{L}{M_t}$ for all $k$, then we have $\|\bc_{\mathrm{opt}}\|_2^2=1$.  It is important to point out that $\bb$ consists of input bits (not output bits) of the convolutional encoder, which results in $B=\frac{B_{in}}{L}$ bits per channel entry quantization.

The procedure of TEC can be summarized as follows: 1)~for a given $\theta$, find the path $\bp_T$ that minimizes the path metric defined in \eqref{TEC_path_metric} by running the Viterbi algorithm; 2) among selected candidate paths depending on $\theta$, select the best path $\bp_{\mathrm{best}}$ that gives the minimum path metric; 3) $\bp_{\mathrm{best}}$ is converted to the binary feedback sequence $\bb$ as in \eqref{c_and_b} and $\bb$ is fed back to the base station; and 4) the base station reconstructs $\bc_{\mathrm{opt}}$ based on $\bb$.

Note that searching over $\theta$ only increases complexity, not the feedback overhead of TEC.  The base station only needs to know the binary feedback sequence $\bb$ that represents the best path $\bp_{\mathrm{best}}$ to reconstruct $\bc_{\mathrm{opt}}$ using the convolutional encoder.  We fix the starting state of the trellis search to the first state.  Otherwise, we need an additional feedback overhead to indicate the starting state of the best path.

\subsection{Codeword-to-branch mapping and codebook design criteria for TEC}\label{cb_design_mapping}
To exploit a pre-existing vector quantized codebook in TEC, we need a clever mapping rule between codewords in $\mathcal{C}^L_{2^{B_{tot}}}$ and branches in the trellis. The mapping rule should depend on the structure of the given trellis or convolutional encoder.  We propose a mapping rule for the trellis structure in Fig. \ref{trellis_rep} for an arbitrary codebook $\mathcal{C}^L_{2^{B_{tot}}}$.  Similar mapping rules can be defined for other trellis structures.

\vspace{0.3cm}
\noindent \textit{1) Codeword-to-branch mapping rule for Fig. \ref{trellis_rep}:}

\begin{figure}[t]
  \centering
  \includegraphics[scale = 0.6]{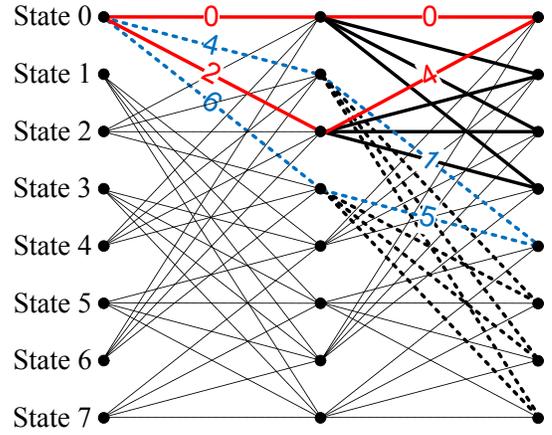}\\
  \caption{Distinctive pairs of paths of which the Euclidean distance should be maximized.  Two pairs of paths are highlighted with trellis outputs.}\label{trellis_dfree}
\end{figure}

Because we fix the starting state of the trellis search as the first state in TEC, we only need to consider the distinctive pairs of paths in Fig. \ref{trellis_dfree}.  Considering the red-solid paths and the first state transition of the blue-dot paths, we can conclude that we need to maximize the minimum Euclidean distance between codeword pairs that are mapped to all even outputs.  For odd outputs, however, we need to separately maximize the Euclidean distance between the two codewords that are mapped to outputs $\left\{1,5\right\}$ and $\left\{3,7\right\}$.

To realize this, with some abuse of notation, let $\mathcal{C}^L_{1}$ and $\mathcal{C}^L_{2}$ denote all possible partitions of $\mathcal{C}^L_{2^{B_{tot}}}$ satisfying
\begin{equation*}
    \mathcal{C}^L_{1} \cup \mathcal{C}^L_{2} = \mathcal{C}^L_{2^{B_{tot}}},
\end{equation*}
\begin{equation*}
    \mathcal{C}^L_{1} \cap \mathcal{C}^L_{2} = \phi,
\end{equation*}
\begin{equation*}
    \mathrm{card}(\mathcal{C}^L_{1}) = \mathrm{card}(\mathcal{C}^L_{2})=2^{B_{tot}-1}
\end{equation*}
where $\mathrm{card}(\cdot)$ is the cardinality of an associated set and $\phi$ denotes an empty set.  Let $\bc_{m,k}\in \mathcal{C}^L_{k}$ for $k=1,2$.  We denote $\mathcal{C}^L_{odd}$ and $\mathcal{C}^L_{even}$ as the set of codewords mapped to the trellis branches of odd and even outputs, respectively.  We generate $\mathcal{C}^L_{{odd}}$ and $\mathcal{C}^L_{{even}}$ as
\begin{align}
\nonumber \mathcal{C}^L_{{odd}}&=\argmax_{\mathcal{C}^L_{1}\subset \mathcal{C}^L}\min_{m\neq n} \left\Vert\bc_{m,1}-\bc_{n,1}\right\Vert^2_2,\\
\mathcal{C}^L_{{even}}&=\argmax_{\mathcal{C}^L_{2}\subset \mathcal{C}^L}\min_{m\neq n} \left\Vert\bc_{m,2}-\bc_{n,2}\right\Vert^2_2, \label{even_C}
\end{align}
respectively.  Once we have $\mathcal{C}^L_{{odd}}$ and $\mathcal{C}^L_{{even}}$ as above, we can have arbitrary mappings between the codewords in $\mathcal{C}^L_{{even}}$ and the trellis branches of even outputs.  For the trellis branches of odd outputs, however, we need one more step.  We divide $\mathcal{C}^L_{odd}$ into $\mathcal{C}^L_{odd,1}$ and $\mathcal{C}^L_{odd,2}$ as we divide $\mathcal{C}^L_{2^{B_{tot}}}$ into $\mathcal{C}^L_{odd}$ and $\mathcal{C}^L_{even}$ in \eqref{even_C}.  Then, we map the codewords in $\mathcal{C}^L_{odd,k}$ to the trellis branches with outputs $\left\{(2k-1),(2k+3)\right\}$ for $k=1,2$.

\vspace{0.3cm}
\noindent \textit{2) Codebook design criterion:}

Instead of reusing conventional codebooks, we can also design a codebook that is optimized for TEC.  Note that the second term of the path metric in \eqref{TEC_path_metric} is the quantization problem in Euclidean space.  Thus, we can generate a codebook with $2^{B_{out}}$ codewords of dimension $L\times 1$ that maximize the minimum Euclidean distance between all possible codeword pairs as
\begin{equation}\label{opt_TEC_codebook}
  \mathcal{C}^L_{\mathrm{ED},2^{B_{out}}}=\argmax_{\mathcal{C}\in\mathcal{U}_{L}^{2^{B_{out}}}}d^2_{ED,\min}(\mathcal{C})
\end{equation}
where $\mathcal{U}_{L}^{N}\in \mathbb{C}^{L \times N}$ is the set of all $L \times N$ complex matrices with unit norm columns and
\begin{equation*}
    d_{ED,
    \min}^2(\mathcal{C})\triangleq \min_{1\leq k<l\leq 2^{N}}\|\bc_k-\bc_l\|_2^2
\end{equation*}
with $\bc_k,\bc_l\in \mathcal{C}$.

The proposed codebook design criterion exploits the same concept as the GLP codebook that maximizes the minimum chordal distance between all codeword pairs \cite{grass1,grass3}.  The difference is that the GLP codebook directly quantizes a channel on the Grassmann manifold while the proposed codebook works in Euclidean space.

\noindent \textbf{Remark:} A similar codebook design and codeword-to-branch mapping criteria have been proposed in \cite{CK}.  However, \cite{CK} first generates the $L \times 1$ Euclidean codebook with $2^{B_{in}}$ codewords (not $2^{B_{out}}$ codewords as in the proposed scheme) that are mapped to odd (or even) outputs.  With some abuse of notation, denote this Euclidean codebook $\mathcal{C}_{odd}^L$.  Then $\mathcal{C}_{even}^L$ is generated by rotating $\mathcal{C}_{odd}^L$ with a unitary matrix $\bU$ where $\bU$ is designed to maximize the minimum chordal distance between codewords in $\mathcal{C}_{odd}^L \cup \mathcal{C}_{even}^L$.  Because $\bU$ tries to maximize the minimum chordal distance, not the minimum Euclidean distance, the approach in \cite{CK} cannot guarantee to maximize the minimum Euclidean distance between all possible pairs of codewords generated by TEC.  Moreover, \cite{CK} cannot utilize an existing vector-quantized codebook different from TEC.

\subsection{TEC for multiple receive antennas}\label{TEC_MIMO}
We can easily modify the proposed TEC to accommodate MIMO with $M_r$ receive antennas at the user.  Assume that $M_t \geq M_r$ and the base station transmits $K\leq M_r$ data streams simultaneously.  Then, we need to quantize the first $K$ dominant eigenvectors of $\bH \bH^H$, which is denoted as $\bU\left(\bH\right) \in \mathbb{C}^{M_t \times K}$.  We can rewrite the path metric defined in \eqref{TEC_path_metric} as
\begin{align*}
&m(\bp_t,\theta)\\
&= \left\Vert \bU\left(\bH\right)_{[1:Lt]}-e^{j\theta}\mathrm{out}(\bp_t)\right\Vert_F^2\\
&= m(\bp_{t-1},\theta) + \left\Vert \bU\left(\bH \right)_{[L(t-1)+1:Lt]}-e^{j\theta}\mathrm{out}([p_{t-1}~p_t])\right\Vert_F^2
\end{align*}
where $\bA_{[m:n]}$ is the truncated matrix of $\bA$ from the $m$th row to the $n$th row, and $\|\bA\|_F$ denotes the Frobenius norm of a matrix $\bA$.

For the multiple receive antenna case, instead of using vector codewords $\bc^L_k$, we need to use matrix codewords $\bC^{L \times K}_k\in \mathbb{C}^{L \times K}$ to quantize $\bU\left(\bH\right)$.  We can use the same codebook design and codeword-to-branch mapping criteria to the multiple receive antenna case by changing the 2-norm operation to a Forbenius norm operation.

\section{Trellis-Extended Successive Phase Adjustment (TE-SPA)}\label{sec4}
In practice, channels are correlated in time and space.  There has been much work on differential codebooks that leverage the temporal correlation of channels for better CSI quantization, e.g., \cite{tm_correlated1,tm_correlated2,tm_correlated7,tm_correlated8,tm_correlated3,tm_correlated4,tm_correlated9,tm_correlated5,tm_correlated6}.  However, most of those works focused on a small number of transmit antennas and feedback bits.  Thus, we first propose TE-SPA which is a differential codebook version of TEC for temporally correlated massive MIMO systems.  Later, we show that TE-SPA can be applied to spatially correlated channels as well.

We consider temporally correlated channels that are modeled by a first order Gauss-Markov process as
\begin{equation}\label{cor_channel_model}
\bh[k] = \eta\bh[k-1] + \sqrt{1-\eta^2} \bg[k]
\end{equation}
where $0\leq \eta \leq 1$, $\bh[k]$, and $\bg[k]$ are the correlation coefficient, the channel realization at time $k$, and the innovation process at time $k$, respectively.  We assume that $\bh[0]$ is independent of $\bg[k]$ for all $k$.  Note that the model in \eqref{cor_channel_model} is also applicable to frequency correlated channels if $k$ denotes the subcarrier or subband index of a wideband channel.

If the channel variation is small in time, i.e., $\eta$ is close to 1, we can successively reduce quantization error by adjusting the phase of each entry or the block of entries of previous CSI.  TE-SPA adjusts phases in a block-wise manner to reduce the feedback overhead. TE-SPA consists of \textit{block-wise phase adjustment matrix generation} and \textit{block shifting}.

\subsection{Block-wise phase adjustment matrix generation}
Let $\hat{\bh}_{k-1} = \bc_{\mathrm{opt}}[k-1]$ and $\bh_k=\bh[k]$ represent the previous (quantized) CSI and the current channel vector, respectively, to simplify notations.  TE-SPA quantizes the channel at time $k$ by adjusting the phases of $\hat{\bh}_{k-1}$ in a block-wise manner.  That is, $\hat{\bh}_{k-1}$ is rotated with a block-wise phase adjustment matrix $\bP_k$ which is given as\footnote{The block length $L$ with the same phase $\varphi_{k,n}$ in $\bP_k$ is a design parameter and does not need to be the same as that of TEC.  We assume the length of $L$ is the same as in TEC for simple explanation.}
\begin{equation}\label{rot_mtx}
\bP_k = \diag\left(\left[e^{j\varphi_{k,1}},\ldots,e^{j\varphi_{k,T}}\right]\otimes\mathbf{1}_L\right)
\end{equation}
where $T=\frac{M_t}{L}$, $\otimes$ is the Kronecker product, and $\mathbf{1}_L=[1,\ldots,1]^T$ is the length $L$ all 1 vector.  Then, the quantized version of the current CSI becomes
\begin{equation*}
\hat{\bh}_{k} = \bP_k\hat{\bh}_{k-1}.
\end{equation*}

TE-SPA exploits the trellis structure as in TEC to generate $\bP_k$, i.e., TE-SPA selects $\varphi_{k,n}$'s from a given set $\Psi=\left\{\psi_1,\ldots,\psi_{2^{B_{out}}}\right\}$ as
\begin{equation}\label{tespa_opt}
\left(\varphi_{k,1},\ldots,\varphi_{k,T}\right) = \argmin_{\varphi_{k,n} \in
\Psi}\min_{\theta\in\Theta}\left\Vert\bh_k-e^{j\theta}\bP_k\hat{\bh}_{k-1}\right\Vert_2^2
\end{equation}
using the Viterbi algorithm.  Note that the convolutional encoders for TEC and TE-SPA can be different, e.g., we could adopt a rate $\frac{2}{3}$ convolutional encoder for TEC while a rate $\frac{1}{2}$ convolutional encoder is used for TE-SPA to reduce successive feedback overhead.

To quantize CSI effectively, we need to appropriately set the values of the elements in $\Psi$ and assign those elements to the trellis branches, which are exactly the same principles as the codebook design and the codeword-to-trellis branch mapping criteria in TEC.  Previous works on differential codebook tried to optimize codebook update methods taking the temporal correlation coefficient $\eta$ into account.  In TE-SPA, this is implicitly handled during the trellis search, i.e., the trellis search selects the best set of phases for $\bP_k$ which rotates the previous CSI ``close'' to the current channel.  Therefore, it is better to have values of elements in $\Psi$ such that they are able to generate various rotation matrices as possible.  Note that $\bP_k$ is determined by the relation among the $\varphi_{k,n}$'s.  If $T=2$, then $\diag([1,e^{j\frac{\pi}{4}}]\otimes\mathbf{1}_L)$ is the same as $\diag([e^{j\frac{7\pi}{4}},1]\otimes\mathbf{1}_L)$ in terms of $\bP_k$. Thus, we can restrict the search space to $[0,\pi]$ and assign the values in $\Psi$ as
\begin{equation*}
  \psi_{\nu} = \frac{\nu-1}{2^{B_{out}}}\pi,\quad \nu=1,\ldots,2^{B_{out}}.
\end{equation*}

Now, we need a mapping rule between $\psi_{\nu}$'s and trellis outputs.  We consider $\psi_{\nu}$'s as PSK constellation points and follow the same mapping rule as in trellis coded modulation (TCM) \cite{Unger}.  That is, we maximize the minimum Euclidean distance among $\psi_{\nu}$'s that are mapped to the branches with the same incoming/outgoing states by mapping $\psi_{\nu}$ to the trellis output $\nu$.

\noindent \textbf{Remark:} We can further reduce the feedback overhead of TE-SPA. Note that we can rewrite $\bP_k$ in \eqref{rot_mtx} as
\begin{equation*}
\bP_k = e^{j\varphi_{k,1}}\diag\left(\left[1,\ldots,e^{j(\varphi_{k,T}-\varphi_{k,1})}\right]\otimes\mathbf{1}_L\right).
\end{equation*}
Let $\breve{\bP}_k=\diag\left(\left[1,\ldots,e^{j(\varphi_{k,T}-\varphi_{k,1})}\right]\otimes\mathbf{1}_L\right)$.  Then, the objective function in \eqref{tespa_opt} can be rewritten as
\begin{equation*}
\left\Vert\bh_k-e^{j(\theta+\varphi_{k,1})}\breve{\bP}_k\hat{\bh}_{k-1}\right\Vert_2^2.
\end{equation*}
Thus, if we appropriately redesign $\Theta$ for the noncoherent search in \eqref{tespa_opt}, we can always fix the first entry of $\breve{\bP}_k$ as~1 and skip (or fix) the first stage of the trellis search which gives a reduced feedback overhead.

\subsection{Block-shifting}
\begin{figure*}[t]
\begin{equation}\label{tespa_opt_rewrite}
\left(\varphi_{k,1},\ldots,\varphi_{k,T}\right) =\argmin_{\varphi_{k,n} \in
\Psi}\min_{\theta\in\Theta}\left\Vert\bh_k\left[\frac{L}{2}(k-1)\right]_c-e^{j\theta}\bP_k\hat{\bh}_{k-1}\left[\frac{L}{2}(k-1)\right]_c\right\Vert^2_2,\quad k\geq 1.
\end{equation}
\hrulefill
\end{figure*}

If we fix the block structure of the phase adjustment matrix $\bP_k$, then the performance can quickly saturate because we cannot adjust the phase relation of the elements within each block.  Moreover, since we fix the starting state of the trellis search, the first state transition suffers from using a restricted number of branches, e.g., only 4 branches with even trellis outputs are exploited for the first state transition in Fig. \ref{trellis_dfree}.  These might not be serious problems for one-shot quantization as in TEC, but the loss could be accumulated in successive quantizations as in TE-SPA.  Therefore, we adopt \textit{block-shifting} to mitigate these problems.
\begin{figure}[t]
  \centering
  \includegraphics[scale = 0.4]{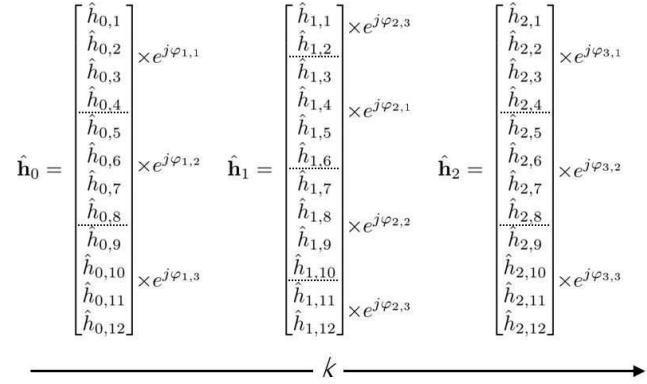}\\
  \caption{A conceptual explanation of TE-SPA with block-shifting with $M_t=12$ and $L=4$.  $\hat{\bh}_k$ is the result of multiplying $e^{j \varphi_{k,n}}$'s to $\hat{\bh}_{k-1}$ in a block-wise manner.}\label{tespa}
\end{figure}

Let $\ba[m]_c$ and $\bA[m]_c$ denote the left circularly shift of a vector $\ba$ and diagonal entries of a matrix $\bA$ of $m$ elements, respectively.  For example, if $\ba=\left[1,2,3,4,5\right]$, then $\ba[2]_c=\left[3,4,5,1,2\right]$.  Using this notation, we rewrite the optimization problem in \eqref{tespa_opt} as in \eqref{tespa_opt_rewrite}.  We interweave two consecutive blocks by circularly shifting $\frac{L}{2}$ elements in \eqref{tespa_opt_rewrite} to prevent the saturation effect.\footnote{To further improve performance, we can dynamically reassign the blocks of $\bP_k$ instead of circularly shifting elements in time.}  After generating $\bP_k$, the quantized CSI at time $k$ is given as
\begin{equation*}
\hat{\bh}_{k} = \bP_k\left[-\frac{L}{2}(k-1)\right]_c\hat{\bh}_{k-1}.
\end{equation*}

The conceptual explanation of TE-SPA with block-shifting is shown in Fig. \ref{tespa}.  Note that TEC is used for CSI quantization at $k=0$.  The proposed block shifting can adjust not only the phase relation among blocks but also that of elements within each block in time.  Moreover, the phase $\varphi_{k,1}$ from the first state transition is multiplied to the different blocks of $\hat{\bh}_k$ depending on $k$, which prevents the accumulation of the loss caused by the first state transition.

\subsection{Applying TE-SPA to spatially correlated channels}\label{te_spa_adap}
In massive MIMO systems, channels tend to be spatially correlated due to small antenna spacing.  We can model spatially correlated channels as
\begin{equation*}
  \bh[k] = \bR^{\frac{1}{2}}\bh_w[k]
\end{equation*}
where $\bR=E[\bh[k]\bh^H[k]]$ is a spatial correlation matrix and $\bh_w[k]$ is uncorrelated channel vector with i.i.d. complex Gaussian entries.  Let $\bu_1(\bR)$ denote the dominant eigenvector of $\bR$.  We assume $\bR$ is perfectly known only at the receive side.

If the channels are highly correlated in space, the matrix $\bR$ becomes ill conditioned, and $\bu_1(\bR)$ and $\bh[k]$ tend to be highly correlated.  In this case, we can quantize $\bu_1(\bR)$ using TEC and apply TE-SPA to quantize $\bh[k]$ in each fading block of $k$ based on the quantized version of $\bu_1(\bR)$.  Because $\bu_1(\bR)$ is a long-term statistic and varies very slowly compared to $\bh_w[k]$, the additional feedback overhead for $\bu_1(\bR)$ would be negligible.  Although this approach is based on one-step (instead of successive) phase adjustment, we keep the terminology TE-SPA to avoid any confusion.

\section{Simulations}\label{sec5}
We performed Monte-Carlo simulations using 10000 channel realizations to evaluate the proposed TEC and TE-SPA.  We set $K_{\theta}=16$ for $\Theta=\left\{\theta_1,\ldots,\theta_{K_{\theta}}\right\}$ to perform the noncoherent search of TEC and TE-SPA.

\begin{figure}[t]
  \centering
  \includegraphics[width=1\columnwidth]{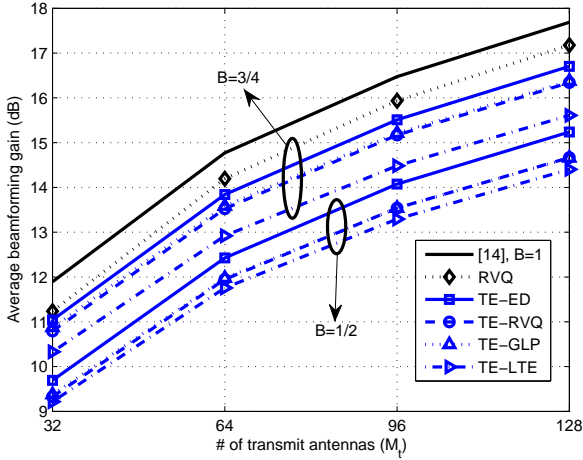}\\
  \caption{Average beamforming gain (dB) with $M_t$ in i.i.d. Rayleigh fading channels.  TE-`codebook name' refers to TEC using the specific codebook.  $B_{tot}=BM_t$.}\label{iid_cb}
\end{figure}

We first evaluate TEC in i.i.d. Rayleigh fading channels as $\bh\sim \mathcal{C}\mathcal{N}(\mathbf{0},\bI_{M_t})$.  Because our focus is CSI quantization techniques, we use the average beamforming gain in dB scale that is defined as
\begin{equation*}
10\log_{10}\left(E[|\bh^H\bc_{\mathrm{opt}}|^2]\right)
\end{equation*}
for a performance metric where the expectation is taken over $\bh$.  We set $L=4$ to exploit vector-quantized codebooks with dimension $4 \times 1$.  Thus, TEC schemes with $B=3/4$ and $B=1/2$ bits per entry quantize 4 channel elements using 3 bits and 2 bits, respectively.  In Fig. \ref{iid_cb}, we plot the average beamforming gain of TEC with the proposed codeword-to-branch mapping rule using different codebooks, e.g., trellis extended-Euclidean distance (TE-ED) refers to TEC using the Euclidean distance (ED) codebook defined in \eqref{opt_TEC_codebook}, according to the number of transmit antennas $M_t$.  We also plot the average beamforming gain of the scheme in \cite{tcom_ntcq} with $B=1$ and that of RVQ with the same feedback overhead with TEC schemes for comparison purpose.  Note that TE-LTE with $B=1/2$ refers to TEC using only the first 8 among 16 codewords of LTE 4 transmit antennas codebook, which are the same as 8 DFT codewords.  The total feedback overhead of each scheme is given as $B_{tot}=BM_t$.
\begin{figure}[t]
  \centering
  \includegraphics[width=1\columnwidth]{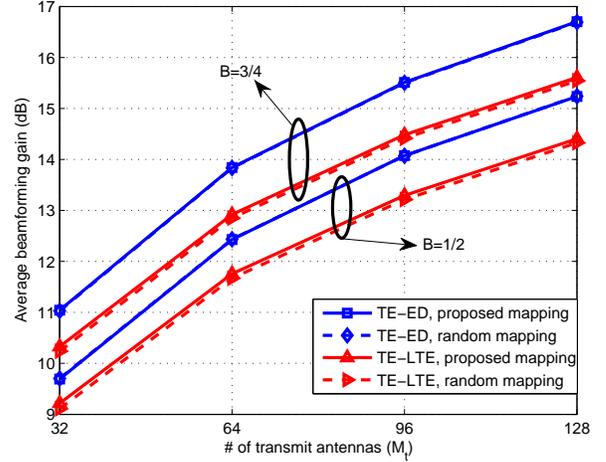}\\
  \caption{Average beamforming gain (dB) with $M_t$ in i.i.d. Rayleigh fading channels.  TEC schemes with the proposed codeword-to-branch mapping and random mapping are compared.}\label{iid_mapping}
\end{figure}

As expected in Section \ref{cb_design_mapping}, TE-ED using the ED codebook gives the best performance among the TEC schemes.  The gain is more than 1 dB compared to TE-LTE when $B=3/4$ and $M_t$ is more than 64.  TE-LTE suffers from practical constraints\footnote{The practical constraints lead to the decreased minimum Euclidean distance among the LTE codewords as well.} on its codewords such as constant modulus (which causes the loss of norm information of channel elements) and finite alphabet properties.  The conventional vector-quantized codebook approach using RVQ is better than TEC schemes, but the plot of the RVQ codebook is based on the analytical approximation of $M_t\left(1-2^{-\frac{B_{tot}}{M_t-1}}\right)$ \cite{rvq3} because it is infeasible to simulate the performance of the RVQ codebook with $B_{tot}=16$ bits (which is the case of $M_t=32$ with $B=1/2$) or more.  The scheme from \cite{tcom_ntcq} outperforms TEC with a much larger feedback overhead than TEC schemes.\footnote{We did not compare TEC with \cite{CK} because the proposed scheme in \cite{CK} cannot even maintain a constant performance gap with the RVQ codebook.}

We also compare the beamforming gains of the proposed codeword-to-branch mapping and a random mapping (per iteration) using TE-ED and TE-LTE in Fig. \ref{iid_mapping}.  Note that the proposed mapping has negligible impact on the average beamforming gain of TE-ED.  The reason is that the Euclidean distance among codewords in the ED codebook is already far apart and the random mapping is also guaranteed to have a good Euclidean distance property.  On the other hand, the proposed mapping achieves around 0.1 to 0.2 dB gain compared to the random mapping in TE-LTE.  This shows that if we reuse pre-existing vector-quantized codebooks that are not optimized in the Euclidean distance, the proposed mapping can achieve additional gain with the same codebook.

\begin{figure}[t]
  \centering
  \includegraphics[width=1\columnwidth]{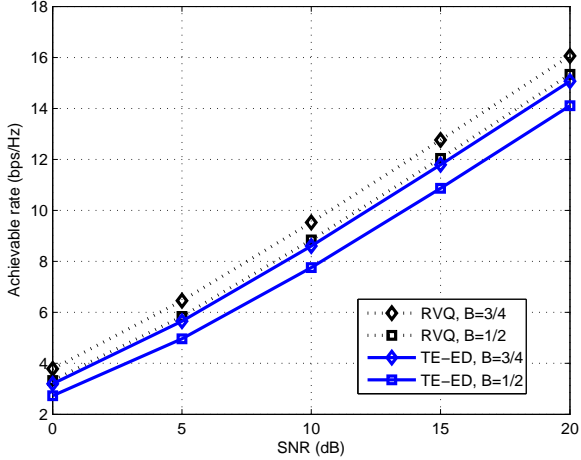}\\
  \caption{Achievable rate with SNR in i.i.d. Rayleigh fading channels with $M_t=16$, $M_r=2$, and $K=2$.  $B_{tot}=B M_t$.}\label{iid_high_rank}
\end{figure}

Now, we evaluate TEC for a multiple receive antenna case.  We set $M_t=16$, $M_r=2$, and the transmission rank as $K=2$.  The number of transmit antennas is \textit{not too large} in this case because we want to compare TEC and the RVQ codebook with the same feedback overhead.  With $M_t=16$, TEC with $B=3/4$ and $B=1/2$ correspond to $B_{tot}=12$ and $B_{tot}=8$ bits, respectively.  Denote the average achievable rate as
\begin{equation*}
  E\left[\log_2\mathrm{det}\left(\bI_{K}+\frac{P}{\sigma^2 K} \bF^H\bH\bH^H \bF \right)\right]
\end{equation*}
where $\frac{P}{\sigma^2}$ is the signal-to-noise ratio (SNR), $\bH\in \mathbb{C}^{M_t \times M_r}$ is the channel matrix, $\bF\in \mathbb{C}^{M_t \times K}$ is the precoder matrix, and the expectation is taken over $\bH$.  Each entry of $\bH$ is distributed with $\cC\cN(0,1)$.  We plot the average achievable rates of TE-ED and RVQ in Fig. \ref{iid_high_rank} with SNR.  The proposed TE-ED maintains a constant gap of around 1 bps/Hz loss compared to the RVQ codebook with the same feedback overhead for all SNR values.  Considering the asymptotic optimality of the RVQ codebook in high rank transmission \cite{rvq2}, the proposed TEC can achieve a good performance even in multiple receive antenna cases with feasible complexity.

\begin{figure}[t]
  \centering
  \includegraphics[width=1\columnwidth]{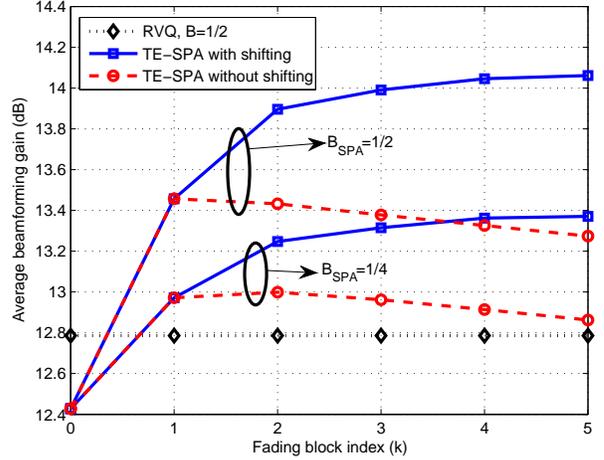}\\
  \caption{Average beamforming gain (dB) with $k$ and $M_t=64$ in temporally correlated channels.  Channels are quantized using TE-ED with $B=1/2$ bits per entry at $k=0$ for TE-SPA schemes.  Total feedback overhead of TE-SPA at $k\geq 1$ is $B_{tot}=B_{SPA}M_t$.}\label{diff}
\end{figure}

In Fig. \ref{diff}, we evaluate TE-SPA with $M_t=64$ in temporally correlated Rayleigh fading channels which is shown in \eqref{cor_channel_model} with $\bg[k]\sim \cC\cN\left(\mathbf{0},\bI_{M_t}\right)$.  We rely on Jakes' model for the temporal correlation coefficient \cite{prok} such that $\eta = J_0(2\pi f_D \tau)$ where $J_0(\cdot)$ is the zero-th order Bessel function, $f_D$ is the maximum Doppler frequency, and $\tau$ is the channel instantiation interval.  With practical system parameters of 2.5GHz carrier frequency, $\tau=5ms$, and $3km/h$ user velocity, the temporal correlation coefficient is given as $\eta=0.9881$.  We do not consider any feedback delay in this simulation because it has been shown in \cite{tcom_ntcq} that the impact of feedback delay is marginal.

At $k=0$, channels are quantized with TE-ED using $B=1/2$ bits per channel entry while channels are quantized using TE-SPA using $B_{SPA}$ bits per entry when $k\geq 1$.  As shown in the figure, the average beamforming gain increases with $k$ due to reduced quantization error using TE-SPA even with lower feedback overhead of $B_{SPA}=1/4$.  All TE-SPA schemes outperform the RVQ codebook that does not consider temporal correlation of channels in quantization.  The gain of using TE-SPA with block shifting is more than 1.6 dB when $B_{SPA}=1/2$.  Note that TE-SPA with block shifting gives far better performance than TE-SPA without block shifting because it can adjust the phase relation of the elements within each block and spread out the loss from the first state transition as explained in Section \ref{sec4}.

\begin{figure}[t]
  \centering
  \includegraphics[width=1\columnwidth]{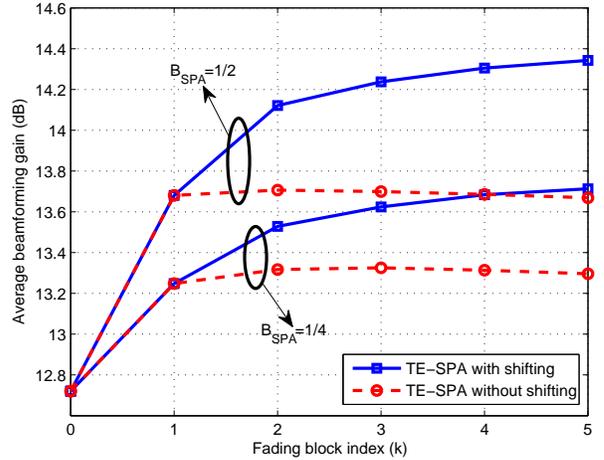}\\
  \caption{Average beamforming gain (dB) with $k$ and $M_t=64$ using an SCM channel model.  Simulation setups are the same as in Fig. \ref{diff} with uniform linear array antennas with $0.5\lambda$ antenna spacing and 8 degrees angle spread.}\label{diff_scm}
\end{figure}

To evaluate TE-SPA in a more practical scenario, we perform simulations using the spatial channel model (SCM) \cite{scm} that is commonly adopted in standards such as 3GPP.  In Fig. \ref{diff_scm}, we plot the average beamforming gain using the same simulation setups as in Fig. \ref{diff} with uniform linear antenna array with $0.5 \lambda$ antenna spacing and 8 degrees angle spread.  As clearly shown in the figure, TE-SPA also works for the practical scenario.

Finally, we evaluate TE-SPA in spatially correlated channels.  We adopt the exponential model \cite{exp_model} for the spatial correlation matrix $\bR$, which is defined as
\begin{align*}
  [\bR]_{\ell,r} = \begin{cases}(\alpha e^{j\vartheta})^{r-\ell},\quad &\ell\leq r \\ [\bR]_{\ell,r}^*,\quad &\ell> r\end{cases},
\end{align*}
where $[\bR]_{\ell,r}$ is the $(\ell,r)$-th element of $\bR$ and $\alpha$ and $\vartheta$ are the magnitude and the phase of the correlation coefficient, respectively.  We set $\alpha=0.9$ to mimic a high spatial correlation of a massive MIMO system while $\vartheta\in [0,2\pi)$ is uniformly randomly generated in each channel realization.

\begin{figure}[t]
  \centering
  \includegraphics[width=1\columnwidth]{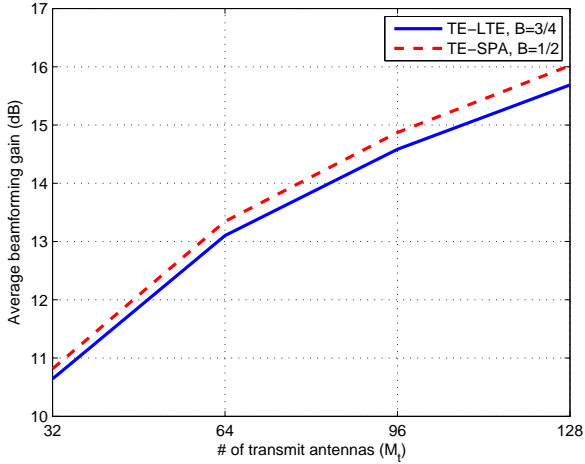}\\
  \caption{Average beamforming gain (dB) with $M_t$ in spatially correlated Rayleigh fading channels.  TE-LTE with $B=3/4$ quantizes $\bh[k]$ directly while TE-SPA refers to the scheme of which $\bu_1(\bR)$ is quantized with TE-LTE with $B=1/2$ and $\bh[k]$ is quantized by TE-SPA with $B=1/2$ based on the quantized $\bu_1(\bR)$.}\label{adap}
\end{figure}

As we can see in Fig. \ref{adap}, TE-SPA is also beneficial for spatially correlated channels even with less feedback overhead than TE-LTE which quantizes $\bh[k]$ directly.  It is important to point out that TE-SPA for spatially correlated channels has additional feedback overhead, i.e., we adopt TE-LTE with $B=1/2$ to quantize $\bu_1(\bR)$.  However, as stated in Section \ref{te_spa_adap}, $\bR$ is a long-term statistic, and the feedback overhead for $\bu_1(\bR)$ would be negligible in long-term sense.

\section{Conclusion}\label{sec6}
We proposed the trellis-extended codebook (TEC) which is an efficient channel quantization technique for FDD massive MIMO systems in this paper.  The proposed TEC exploits a trellis quantizer combined with vector-quantized codebooks to achieve a practical feedback overhead and complexity.  TEC can easily satisfy backward compatibility by exploting standardized codebooks such as LTE or LTE-Advanced codebooks.  We proposed a codeword-to-branch mapping and codebook design criteria to maximize the performance of TEC.  TEC also can support multiple receive antennas making a unified CSI quantization framework possible.  It has been shown using simulations that TEC can maintain a constant performance gap with RVQ which is known to be asymptotically optimal.

We also developed a differential codebook version of TEC called trellis-extended successive phase adjustment (TE-SPA).  We incorporated a trellis structure to quantize temporally correlated channels in a successive manner.  TE-SPA also can be adapted to spatially correlated channels without any difficulty.  TEC and TE-SPA can be thought of as an evolution of the LTE-Advanced dual codebooks for long-term/wideband and short-term/subband CSI quantization.  The numerical results confirmed that the proposed TE-SPA can reduce quantization loss even with reduced feedback overhead.

Because TEC and TE-SPA both support various numbers of CSI quantization bits, the proposed techniques can easily allocate different numbers of feedback bits per user based on system requirements or channel conditions \cite{bit_alloc1,bit_alloc2,bit_alloc3}.  This is also a strong benefit for FDD massive MIMO systems of which the feedback overhead needs to be carefully optimized.

\section*{Acknowledgment}
This work was sponsored by Communications Research Team (CRT), DMC R\&D Center, Samsung Electronics Co. Ltd.

\bibliographystyle{IEEEtran}
\bibliography{ref}

\end{document}